\documentstyle[multicol,prl,aps,psfig]{revtex}
\begin{document}
\title{Transport across a normal-superconducting interface: a novel probe of
electron-electron interactions in the normal metal.}
\author{P. Dolby, R. Seviour and C.J.Lambert}
\address{Department of Physics, Lancaster University,
 Lancaster LA1 4YB, UK.}
\maketitle
\begin{abstract}
In this letter, we examine the effect of Coulomb interactions
 in the normal region of a normal-superconducting (N/S)  mesoscopic structure, 
here the change from an attractive to a repulsive coulombic interaction,  
at the $N/S$ interface, causes a shift in the order parameter phase. 
We show that this shift has a pronounced effect on 
Andreev bound states and demonstrate that the effect 
on Andreev scattering of non-zero order-parameter tails, can be used to probe the sign of 
the interaction in the normal region.  
\end{abstract}
\begin{multicols}{2}
\narrowtext
\noindent

 Recent advances in material technology have enabled the fabrication
of  normal/superconducting (N/S) mesoscopic hybrid structures with well defined
dimensions and interfaces \cite{afv:r2,afv:r3,afv:r4,afv:r5}. 
Due to the proximity of the normal material to the 
superconductor the pairing field  
$f(x) = \langle \psi_\uparrow(x) \psi_\downarrow(x) \rangle$ 
, in the normal region, decays to zero on the
scale of a coherence length $\xi$ \cite{rb}. During the past decade 
this proximity effect 
has been extensively investigated both experimentally and theoretically 
 (see for example \cite{kastalsky91,vol1,r2,r1}). 
 
\begin{figure}
\begin{center}
\psfig{figure=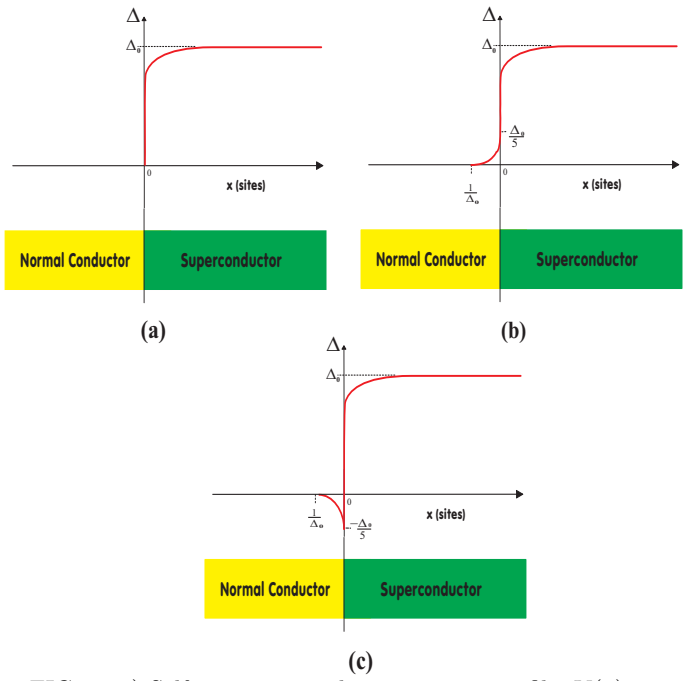,width=9cm,height=9cm}
\caption{a) Self-consistent order parameter profile, V(x)=0 in the normal 
region, b) Self-consistent order parameter profile, V(x) and f(x) both finite 
in normal region, c) Self-consistent order parameter profile including Coulomb 
interactions.}
\label{Fig.1}
\end{center}
\end{figure}

In contrast to the pairing field $f(x)$, the effective electron-electron
 interaction, $V(x)$,
changes abruptly at the S/N interface,
 from an attractive interaction in the superconductor to either zero, 
a much diminished
attractive interaction or to a repulsive interaction, in the normal (N) material.  
Consequently the order parameter, $\Delta = V(x)f(x)$,
of a s-wave superconductor also changes abruptly at the $S/N$ interface, 
as shown in figure \ref{Fig.1}. To date theoretical research into
the transport properties of $N/S$ interfaces have mainly 
considered the order parameter in the normal region to be zero
 (figure \ref{Fig.1}a) \cite{rm,ra}.

 In this letter we consider the effect of an attractive or repulsive electron-electron
interaction ($V(x) \ne 0$), as shown in \ref{Fig.1}b and \ref{Fig.1}c. The difference 
between figures \ref{Fig.1}b and \ref{Fig.1}c is the $\pi$ phase shift in 
$\Delta(x)$, induced by the cross-over from an attractive to a repulsive 
interaction at the $N/S$ interface.  In what follows we examine how this phase 
shift affects the transport properties associated with Andreev bound states.
  
\begin{figure}[ht]
\begin{center}
\psfig{figure=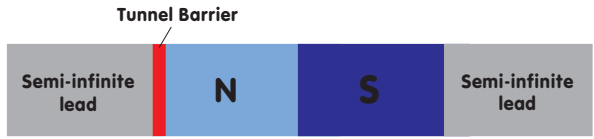,width=9cm,height=3cm}
\caption{A N-S system with a tunnel barrier present.  Such a system will have
Andreev bound states present in the region between the interface and the tunnel
barrier.}
\label{Fig.2}
\end{center}
\end{figure}

To investigate this regime we adopt a general scattering approach 
to dc transport, which was initially developed to describe phase-coherent 
transport in dirty mesoscopic superconductors \cite{c17}. For simplicity 
in this letter, we focus solely on the zero-voltage, zero-temperature conductance, 
for the structure shown in figure \ref{Fig.2}.  In the linear-response limit, at zero temperature, the conductance of a
 phase-coherent structure may be calculated from the fundamental current 
voltage relationship \cite{c18,c19},
\begin{equation} 
I_i=\sum_{j=1}^2a_{ij}( v_j - v),
\label{a4}
\end{equation}

The above expression relates the current $ I_i$ from a normal reservoir 
$i$ to the voltage differences $(v_j - v)$, where $v=\mu/e$
and the sum is over the 2 normal leads connected to the scattering region.
The $a_{ij}$'s are linear combinations of the normal and Andreev scattering
coefficients and in the absence of superconductivity satisfy $\sum_{j=1}^2a_{ij}
=\sum_{i=1}^2a_{ij} = 0 $ in which case the left hand side of equation \ref{a4}
becomes independent of $v$. In units of
$2e^2/h$ \cite{c18,c19}, $a_{ii}=N_i+R_i^A-R_i^O$ and $a_{ij\ne i}=T_{ij}^A-T_{ij}^O$,
where $T_{ij}^A$, $T_{ij}^O$ are Andreev and normal transmission coefficients from
probe $j$ to probe $i$, $R_i^A$, $R_i^O$ are Andreev and normal reflection
coefficients from probe $i$ and $N_i$ is the number of open scattering
channels  in lead $i$.

 Setting $I_1 = -I_2 = I$ and solving 
equation \ref{a4} for the 2 probe conductance yields \cite{c20},

\begin{equation} 
G=\frac{I}{(V_{1} - V_{2})} =
T^{o}_{21}+T_{12}^{A} + {{2(R_{2}^{A} R_{1}^{A} -T_{21}^{A} T_{12}^{A})}
\over {R_{2}^{A}+R_{1}^{A}+T_{21}^{A}+T_{12}^{A}}},
\label{ce7}
\end{equation}

 where $G$ is the conductance in units of $\frac{2e^{2}}{h}$.
 As noted in \cite{c20} the various transmission 
and reflection coeffcients can be computed by solving the
Bogoliubov - de Gennes equation on a tight-binding lattice of
sites, where each site is labelled by an index $i$ and possess
a particle (hole) degree of freedom $\psi(i)$ $(\phi(i))$.
In the presence of local s-wave pairing described by a
superconducting order parameter $\Delta_i$, this takes the form,

\begin{eqnarray}
\begin{array}{c c}
E\psi_i
=&\epsilon_i \psi_{i}
-\sum_{\delta} \gamma \left(  \psi_{i+\delta} + \psi_{i-\delta} \right)
+ \Delta_{i} \phi_{i}\\
E\phi_i =&-
\epsilon_i \phi_{i}
+\sum_{\delta} \gamma  \left(  \phi_{i+\delta}+ \phi_{i-\delta}\right)
+\Delta^*_{i}\psi_{i},\\
\end{array}
\label{2}
\end{eqnarray}

In what follows, the Hamiltonian of eq.(\ref{2}) is used to describe 
the structure of figure \ref{Fig.2}, where for all i, the on-site energy $\epsilon_i=\epsilon_0$. In the S-region, the 
order parameter $\Delta_i$ is set to a constant, $\Delta_i=\Delta_0$, while in the 
normal region $\Delta_j$ is approximated by,

\begin{equation} 
\Delta_j= \pm\frac{\Delta_0}{5} \left ( \tanh (j-L_n) +1 \right ).
\label{app}
\end{equation}

 The nearest neighbour hopping element $\gamma$ merely 
fixes the energy scale (ie the band-width), whereas $\epsilon_0$ determines the
band-filling, and  $L_n$ is the length of the
normal region. In what follows we choose $\gamma=1$ and $\Delta_0 = 0.1$.
By numerically solving for the scattering matrix of equation \ref{2},
exact results for the dc conductance can be
obtained and therefore the effects of a repulsive/attractive coulomb 
interaction in the normal region can be examined.

\begin{figure}
\centerline{\psfig{figure=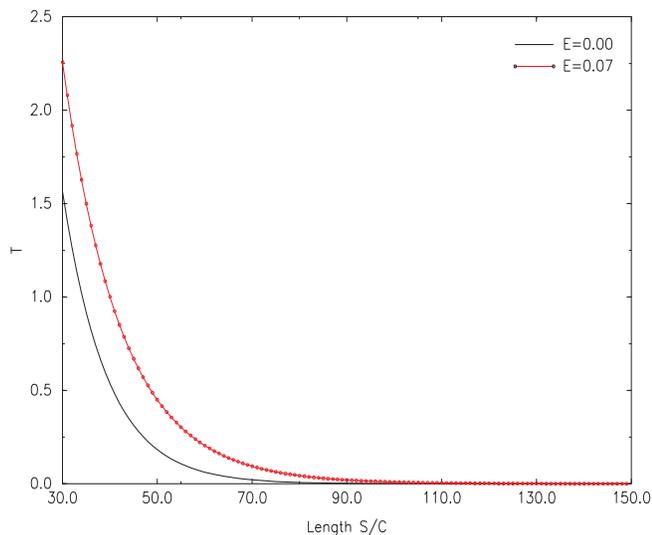,width=8.5cm,height=7cm}}
\caption{Graph showing the transmission coeffcient as a function of Superconductor length for
various quasiparticle energies.  No Barrier or proximity tail are present. 
Structure width is 20 sites.  In this graph it can be seen that at high quasiparticle energies 
transmission will occur through the superconductor.}
\label{Fig.3}
\end{figure}

\begin{figure}
\centerline{\psfig{figure=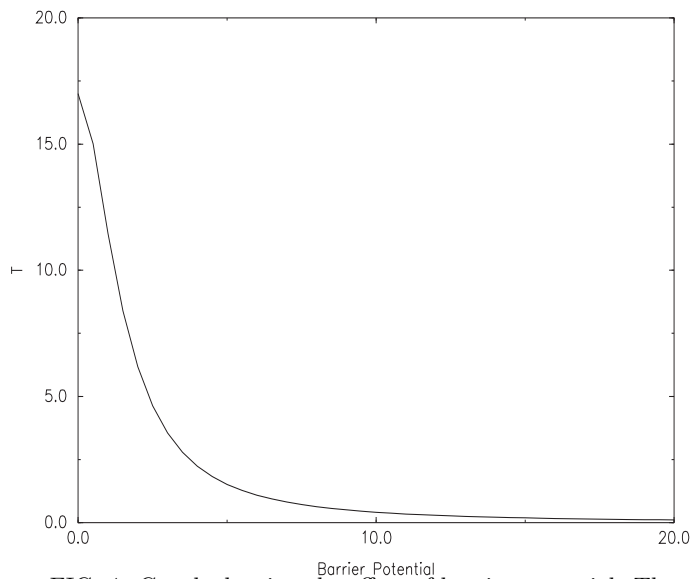,width=9cm,height=7.5cm}}
\caption{Graph showing the effect of barrier potential.  The barrier is one site
wide, structure width is 20 sites.  The graph shows transmission as a function of barrier potential for a quasiparticle
energy, E=0.00.}
\label{Fig.4}
\end{figure}

  For simplicity in this letter we consider the transport properties of the 
 structure shown in figure \ref{Fig.2}.  
 Consisting of two normal, semi-infinite, crystalline leads, 20 sites wide, 
joined by a scattering region. The normal scattering 
region is 40 sites long and the superconducting region is of length $L_S$.
To form Andreev bound states we create quasiparticle confinement in the 
area in front of the superconductor by introducing a weak 
point-like contact between the left lead and the scattering region, and 
ensuring that there is no or very little quasiparticle transmission 
through the superconductor. For this reason the superconductor length was 
set to, $L_S= 150$ sites.  However care must be taken since the decay 
length of sub-gap states in the 
superconductor increases with quasiparticle energy.  At energies close to 
the gap energy the decay length is long enough for transmission through 
the superconductor, see figure \ref{Fig.3}.
 The weak contact is created  by placing a 
potential barrier, one site wide, between 
the left lead and the normal scattering region, via a mean potential $U$ added to 
the on site energy $\epsilon_0$.
 Figure \ref{Fig.4} shows a plot of the transmission coefficient of the barrier
 as a function of $U$. From this plot we see that a for low 
quasiparticle transmission through the barrier a high  potential is needed. 
For these reasons the barrier potential was set at $U=20$.

\begin{figure}
\centerline{\psfig{figure=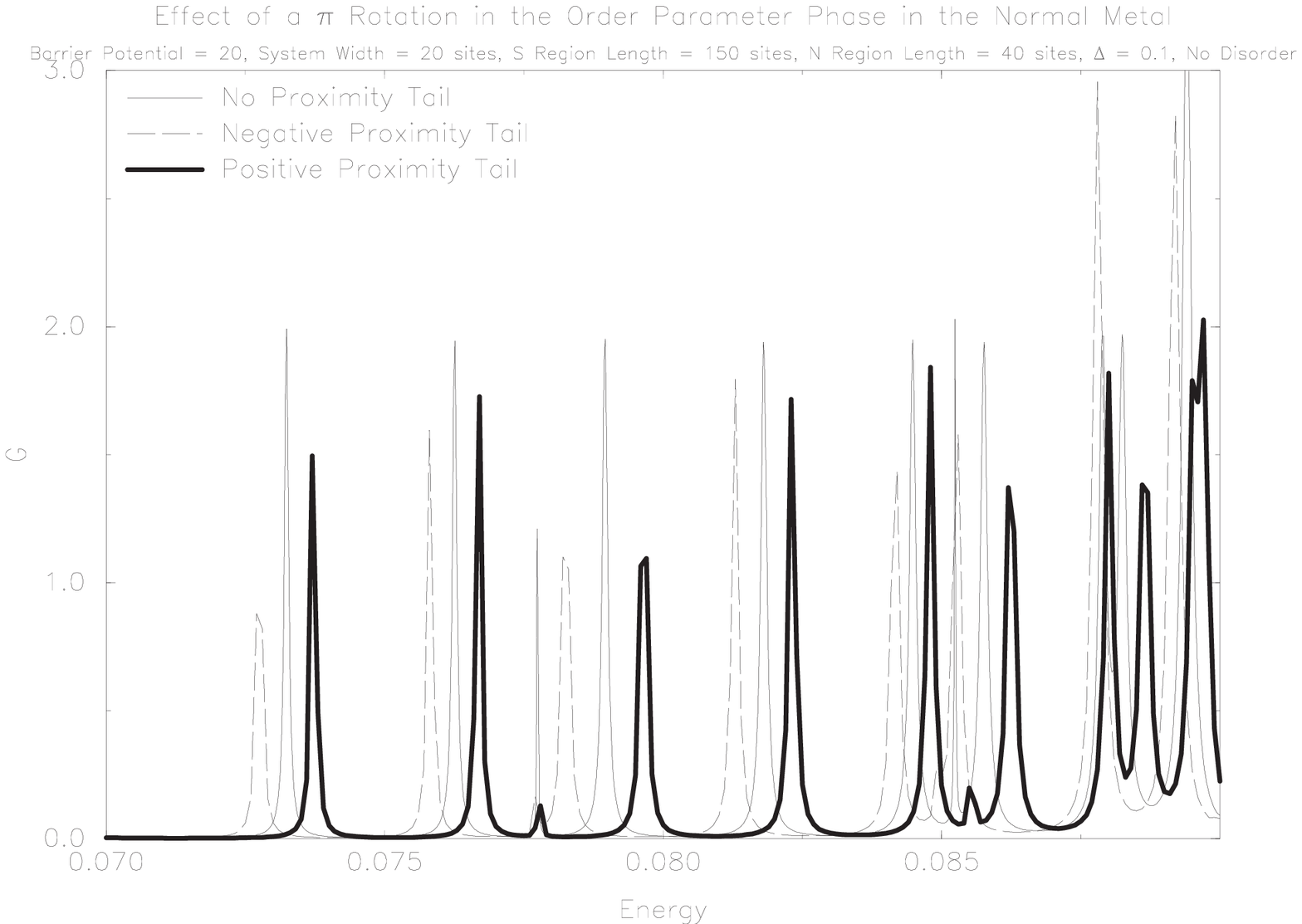,width=9cm,height=7.5cm}}
\caption{Graphs of Conductance as a function of quasiparticle energy for different phases of the order parameter tail in the
Normal metal.  }
\label{Fig.5}
\end{figure}

 These conditions produce discrete states within the normal regions. 
Allowing the formation of Andreev bound states, when these energy levels 
become populated reasonances appear in the conductance of the 
system, analogous to Breit-Wigner resonances\cite{rr}.
 By plotting $G$ as a function of quasiparticle energy (all energies are 
with respect to the Fermi energy) we are able to investigate transport
resonances in the conductance due to the formation of Andreev 
bound states.
Figure \ref{Fig.5} shows the conductance as a function of quasiparticle energy for
the three proximity tails corresponding to the vanishing, attractive and
repulsive electron-electron interactions respectively.  These represent the central result of this letter.
We see that the introduction of a proximity tail has the effect of shifting the energy at 
which the Andreev bound states occur.  A positive proximity tail causes a shift 
in the resonance energies to the right of the zero-tail spectrum shown in figure 
\ref{Fig.5}, whereas a negative proximity tail shifts the spectrum to the left.
These results suggest a novel method for detecting the 
sign of the electron-electron interaction in the N-metal. Namely by suppressing the order parameter in the normal region, 
the resonances will either shift to higher energies indicating a repulsive interaction, or 
lower energies in the case of an attractive interaction. This suppression can be
achieved by, for example applying a magnetic field or via a control current in
the superconductor.

This work is funded by the EPSRC and the E.U. TMR
programme.

\end{multicols}

\begin{references}

\bibitem{afv:r2} V.T. Petrashov, V.N. Antonov, P. Delsing, and T. Claeson,
             Phys. Rev. Lett.{\bf 70}, 347, (1993);
             Phys. Rev. Lett. {\bf 74},5268 (1995).

\bibitem{afv:r3} H. Pothier, S. Gueron, D. Esteve, and M.H. Devoret,
             Phys. Rev. Lett. {\bf 73}, 2488 (1994).

\bibitem{afv:r4} P.G.N. Vegvar, T.A. Fulton, W.H. Mallison, and R.E. Miller,
             Phys. Rev. Lett. {\bf 73}, 1416(1994).

\bibitem{afv:r5} H. Courtois, Ph. Grandit, D. Mailly, and B. Pannetier,
             Phys. Rev. Lett. {\bf 76}, 130 (1996);
             D. Charlat, H. Courtais, Ph. Grandit, D. Mailly, A.F. Volkov,
             and B. Pannetier,
             Phys. Rev. Lett. {\bf 77}, 4950 (1996).

\bibitem{rb}  P. G. de Gennes, Superconductivity of Metals and Alloys (Benjamin,
New York, 1966).

\bibitem{kastalsky91}  Kastalsky A,  Kleinsasser A W,  Greene L H,
 Milliken F P and  Harbison J P 1991  Phys. Rev. Lett. {\bf 67} 3026

\bibitem{vol1} Volkov A.F., Seviour R., Pavlovskii V.V., 
Superlattices and Microstructures, 1999, vol. 25, no. 56, pp. 647-657


\bibitem{r2}  Y. Levi, O. Millo, N. D. Rizzo, D. E. Prober, and L. R. Motowidlo,
Phys. Rev.\ {\bf B58},\ 15128\ (1998)

\bibitem{r1}  Jian-Xin Zhu, C. S. Ting, Phys. Rev.\ {\bf B61},\ 1456\ (2000)

\bibitem{rm}  W. L. McMillan, Phys. Rev. \ {\bf 175},\ 559\ (1968).
\bibitem{ra}  A. M. Martin and James F. Annett, Phys. Rev.\ {\bf B57},\ 8709\ (1998).

\bibitem{c17}   Lambert C J and  Hui V C 
1990 Physica B {\bf 165}  1107 

\bibitem{c18} C.J. Lambert, J. Phys.: Condensed Matter {\bf 3},
 6579 (1991).

\bibitem{c19} C.J. Lambert, V.C. Hui,
and S.J. Robinson, J.Phys.: Condens. Matter,
{\bf 5}, 4187 (1993).

\bibitem{c20} N Allsopp et al, J. Phys. Condens. Matter 10475 {\bf 6} (1994)

\bibitem{rc}  C. J. Lambert and R. Raimondi, J. Phys. Condens. Matter\ {\bf 10},\ 901\ (1998).
\bibitem{rr}  N. R. Claughton, M. Leadbeater and C. J. Lambert, J. Phys. Condens. Matter\ {\bf 7},\ 8757\ (1995). 
\end{references}
\end{document}